\documentclass[epj]{svjour}
\usepackage{graphics}
\usepackage{epsf}
\begin{document}

\def\lesssim{\raise0.3ex\hbox{$\;<$\kern-0.75em\raise-1.1ex\hbox{$\sim\;$}}}
\def\gtrsim{\raise0.3ex\hbox{$\;>$\kern-0.75em\raise-1.1ex\hbox{$\sim\;$}}}

\title{Cosmology and neutrino masses - an update}
\author{Steen Hannestad 
\thanks{hannestad@fysik.sdu.dk}%
}                     
\offprints{}          
\institute{Department of Physics, University of Southern Denmark,
Campusvej 55, DK-5230 Odense M, Denmark}
%
%
\abstract{
Present cosmological observations yield an upper bound on the
neutrino mass which is significantly stronger than laboratory
bounds. However, the exact value of the cosmological
bound is model dependent and therefore less robust. Here, I review
the current status of cosmological neutrino mass bounds and also
discuss implications for sterile neutrinos and LSND in particular.}

%
%
\maketitle
\section{Introduction}
\label{intro}

The absolute value of neutrino masses are very difficult to measure
experimentally. On the other hand, mass differences between neutrino
mass eigenstates, $(m_1,m_2,m_3)$, 
can be measured in neutrino oscillation experiments.

The combination of all currently available data suggests two important
mass differences in the neutrino mass hierarchy. 
The solar mass difference of
$\delta m_{12}^2 \simeq 7 \times 10^{-5}$ eV$^2$ and
the atmospheric mass difference
$\delta m_{23}^2 \simeq 2.6 \times 10^{-3}$ eV$^2$
\cite{Maltoni:2003da,Aliani:2003ns,deHolanda:2003nj}
(see also the contribution by C. Giunti to the present volume).

In the simplest case where neutrino masses are
hierarchical these results suggest that $m_1 \sim 0$, $m_2 \sim 
\delta m_{\rm solar}$, and $m_3 \sim \delta m_{\rm atmospheric}$.
If the hierarchy is inverted 
\cite{Kostelecky:1993dm,Fuller:1995tz,Caldwell:1995vi,Bilenky:1996cb,King:2000ce,He:2002rv}
one instead finds
$m_3 \sim 0$, $m_2 \sim \delta m_{\rm atmospheric}$, and 
$m_1 \sim \delta m_{\rm atmospheric}$.
However, it is also possible that neutrino
masses are degenerate
\cite{Ioannisian:1994nx,Bamert:vc,Mohapatra:1994bg,Minakata:1996vs,Vissani:1997pa,Minakata:1997ja,Ellis:1999my,Casas:1999tp,Casas:1999ac,Ma:1999xq,Adhikari:2000as}, 
$m_1 \sim m_2 \sim m_3 \gg \delta m_{\rm atmospheric}$, 
in which case oscillation experiments are not
useful for determining the absolute mass scale.

Experiments which rely on kinematical effects of the neutrino mass
offer the strongest probe of this overall mass scale. Tritium decay
measurements have been able to put an upper limit on the electron
neutrino mass of 2.2-2.3 eV (95\% conf.) \cite{Bonn:tw}
(see also the contribution by C. Kraus in the present volume).
However, cosmology at present yields an even stronger limit which
is also based on the kinematics of neutrino mass.

Neutrinos decouple at a temperature of 1-2 MeV in the early universe,
shortly before electron-positron annihilation.
Therefore their temperature is lower than the photon temperature
by a factor $(4/11)^{1/3}$. This again means that the total neutrino
number density is related to the photon number density by
\begin{equation}
n_{\nu} = \frac{9}{11} n_\gamma
\end{equation}

Massive neutrinos with masses $m \gg T_0 \sim 2.4 \times 10^{-4}$ eV
are non-relativistic at present and therefore contribute to the
cosmological matter density \cite{Hannestad:1995rs,Dolgov:1997mb,Mangano:2001iu}
\begin{equation}
\Omega_\nu h^2 = \frac{\sum m_\nu}{92.5 \,\, {\rm eV}},
\end{equation}
calculated for a present day photon temperature $T_0 = 2.728$K. Here,
$\sum m_\nu = m_1+m_2+m_3$.
However, because they are so light
these neutrinos free stream on a scale of roughly 
$k \simeq 0.03 m_{\rm eV} \Omega_m^{1/2} \, h \,\, {\rm Mpc}^{-1}$
\cite{dzs,Doroshkevich:tq,Hu:1997mj}. 
Below this scale neutrino perturbations are completely erased and 
therefore the matter power spectrum is suppressed, roughly by
$\Delta P/P \sim -8 \Omega_\nu/\Omega_m$ \cite{Hu:1997mj}.

This power spectrum suppression allows for a determination of the
neutrino mass from measurements of the matter power spectrum on
large scales. This matter spectrum is related to the galaxy correlation
spectrum measured in large scale structure (LSS) surveys via the
bias parameter, $b^2 \equiv P_g(k)/P_m(k)$.
Such analyses have been performed several times before
\cite{Croft:1999mm,Fukugita:1999as}, most recently
using data from the 2dF galaxy survey \cite{Elgaroy:2002bi}. 

However, using large scale structure data alone does not allow for a
precise determination of neutrino masses, because the power spectrum
suppression can also be caused by changes in other parameters, such as
the matter density or the Hubble parameter.

Therefore it is necessary to add information on other parameters
from the cosmic microwave background (CMB). This has been done
in the past \cite{Elgaroy:2002bi,Hannestad:2002xv,Lewis:2002ah},
using ealier CMB data. More recently the precise data from WMAP
\cite{map1}
has been used for this purpose \cite{map2,steen03,el03} to derive
a limit of 0.7-1.0 eV for the sum of neutrino masses.

\section{Cosmological data and likelihood analysis}
\label{sec:1}

The extraction of cosmological parameters from cosmological data
is a difficult process 
since for both CMB and LSS the power spectra depend on a plethora
of different parameters.
Furthermore, since the CMB and matter power spectra
depend on many different parameters one might
worry that an analysis which is too restricted in parameter space 
could give spuriously strong limits on a given parameter.

The most recent cosmological data is in excellent agreement with 
a flat $\Lambda$CDM model, the only non-standard feature being
the apparently very high optical depth to reionization. Therefore
the natural benchmark against which non-standard neutrino physics can
be tested is a model with the following free parameters:
$\Omega_m$, the matter density,
the curvature parameter, $\Omega_b$, the baryon density, $H_0$, the
Hubble parameter, $n_s$, the scalar spectral index of the primordial
fluctuation spectrum, $\tau$, the optical depth to reionization,
$Q$, the normalization of the CMB power spectrum, $b$, the 
bias parameter, and finally the two parameters related to neutrino physics,
$\Omega_\nu h^2$ and $N_\nu$.
The analysis can be restricted to geometrically flat models, i.e.\
$\Omega = \Omega_m + \Omega_\Lambda = 1$.
For the purpose of actual power spectrum calculations, the CMBFAST
package \cite{CMBFAST} can be used.

\subsection{LSS data}

At present, by far the largest survey available
is the 2dFGRS \cite{2dFGRS} of which about 147,000 galaxies have so far been
analyzed. Tegmark, Hamilton and Xu \cite{THX} have calculated a power
spectrum, $P(k)$, from this data, which we use in the present work.
The 2dFGRS data extends to very small scales where there are large
effects of non-linearity. Since we only calculate linear power
spectra, we use (in accordance with standard procedure) only data on
scales larger than $k = 0.2 h \,\, {\rm Mpc}^{-1}$, where effects of
non-linearity should be minimal \cite{Hannestad:2002cn}. 
Making this cut reduces the number of power spectrum data points to 18.

\subsection{CMB data}

The CMB temperature
fluctuations are conveniently described in terms of the
spherical harmonics power spectrum
$C_l \equiv \langle |a_{lm}|^2 \rangle$,
where
$\frac{\Delta T}{T} (\theta,\phi) = \sum_{lm} a_{lm}Y_{lm}(\theta,\phi)$.
Since Thomson scattering polarizes light there are also power spectra
coming from the polarization. The polarization can be
divided into a curl-free $(E)$ and a curl $(B)$ component, yielding
four independent power spectra: $C_{T,l}, C_{E,l}, C_{B,l}$ and 
the temperature $E$-polarization cross-correlation $C_{TE,l}$.

The WMAP experiment have reported data only on $C_{T,l}$ and $C_{TE,l}$,
as described in Ref.~\cite{map1,map2,map3,map4,map5}

We have performed the likelihood analysis using the prescription
given by the WMAP collaboration which includes the correlation
between different $C_l$'s \cite{map1,map2,map3,map4,map5}. Foreground contamination has
already been subtracted from their published data.

In parts of the data analysis we also add other CMB data from
the compilation by Wang {\it et al.} \cite{wang3}
which includes data at high $l$.
Altogether this data set has 28 data points.

\section{Neutrino mass bounds}
\label{sec:2}

The analysis presented here was originally published in Ref.~\cite{steen03},
and more details can be found there.

We have calculated $\chi^2$ as a function of neutrino mass while
marginalizing over all other cosmological parameters. This has been
done using the data sets described above. In the first case we have
calculated the constraint using the WMAP $C_{T,l}$ combined with
the 2dFGRS data, and in the second case we have added the polarization
measurement from WMAP.
Finally we have added the additional constraint from the
HST key project and the Supernova Cosmology Project.
It should also be noted that when constraining the neutrino mass
it has in all cases been assumed that $N_\nu$ is equal to the
standard model value of 3.04. Later we relax this
condition in order to study the LSND bound.

The result is shown in Fig.~1. As can be seen from the figure
the 95\% confidence upper limit on the sum of neutrino masses is
$\sum m_\nu \leq 1.01$ eV (95\% conf.) using the case with priors.
This value is completely consistent with what is found in
Ref.~\cite{el03} where simple Gaussian priors from WMAP were
added to the 2dFGRS data analysis.
For the three cases studied the upper limits on $\sum m_\nu$
can be found in Table 1.

%
\begin{table}
\caption{95\% C.L. upper limits on $\sum m_\nu$ for the three different 
cases: 1) WMAP+Wang+2dFGRS+HST+SN-Ia, 2) WMAP+Wang+2dFGRS
3) WMAP+2dFGRS.}
\label{tab:1}       
\begin{center}
\begin{tabular}{cc}
\hline\noalign{\smallskip}
Case &  $\sum m_\nu$ (95\% C.L.) \\
\noalign{\smallskip}\hline\noalign{\smallskip}
1 & 1.01 eV \\
2 & 1.20 eV  \\
3 & 2.12 eV \\
\noalign{\smallskip}\hline
\end{tabular}
\end{center}
\vspace*{1cm}  
\end{table}

In the middle panel of Fig.~1 we show the best fit value of $H_0$
for a given $\Omega_\nu h^2$. It is clear that an increasing value
of $\sum m_\nu$ can be compensated by a decrease in $H_0$. Even though
the data yields a strong constraint on $\Omega_m h^2$ there is no
independent constraint on $\Omega_m$ in itself. Therefore, an decreasing
$H_0$ leads to an increasing $\Omega_m$. This can be seen in the bottom
panel of Fig.~1.

When the HST prior on $H_0$ is relaxed a higher value of $\sum m_\nu$
is allowed, in the case with only WMAP and 2dFGRS data the upper bound
is $\Omega_\nu h^2 \leq 0.023$ (95\% conf.), corresponding to a
neutrino mass of 0.71 eV for each of the three neutrinos.

This effect was also found by Elgar{\o}y and Lahav \cite{el03} in 
their analysis of the effects of priors on the determination 
of $\sum m_\nu$.

However, as can also be seen from the figure, the addition of high-$l$
CMB data from the Want {\it et al.} compilation also shrinks the
allowed range of $\sum m_\nu$ significantly. The reason is that
there is a significant overlap of the scales probed by high-$l$ CMB
experiments and the 2dFGRS survey. Therefore, even though we use bias
as a free fitting parameter, it is strongly constrained by the fact
that the CMB and 2dFGRS data essentially cover much of the same
range in $k$-space.

It should be noted that Elgar{\o}y and Lahav \cite{el03} find that bias
does not play any role in determining the bound on $\sum m_\nu$. At first
this seems to contradict the discussion here, and also what was found
from a Fisher matrix analysis in Ref.~\cite{Hannestad:2002xv}. The reason
is that in Ref.~\cite{el03}, redshift distortions are included in
the 2dFGRS data analysis. Given a constraint on the amplitude of fluctuations
from CMB data, and a constraint on $\Omega_m h^2$ , this effectively
constrains the bias parameter. Therefore adding a further constraint
on bias in their analysis does not change the results.

{\it Neutrinoless double beta decay --}
Recently it was claimed that the Heidelberg-Moscow experiment
yields positive evidence for neutrinoless double beta decay.
Such experiments probe the `effective electron neutrino mass
$m_{ee} = |\sum_j U^2_{ej} m_{\nu_j}|$. Given the uncertainties in
the involved nuclear matrix elements the Heidelberg-Moscow
result leads to a mass of $m_{ee} = 0.3-1.4$ eV. If this is
true then the mass eigenstates are necessarily degenerate, and
$\sum m_\nu \simeq 3 m_{ee}$.
Taking the WMAP result of $\sum m_\nu \leq 0.70$ eV at face
value seems to 
be inconsistent with the Heidelberg-Moscow result
\cite{Pierce:2003uh}.
However, already if Ly-$\alpha$ forest data and a constraint
on the bias parameter is not used in the analysis
the upper bound of $\sum m_\nu \leq 1.01$ eV is still
consistent.
For this reason it is probably premature to rule out the
claimed evidence for neutrinoless double beta decay.

{\it Evidence for a non-zero neutrino mass --}
In a recent paper \cite{Allen:2003pt} it was noted that there is 
a preference for a non-zero neutrino mass if a measurement of the
bias parameter from X-ray clusters is added to the CMB and large
scale structure data. This result arises because the X-ray data 
prefers a low value of $\sigma_8$ (bias), which is incompatible 
with the WMAP and 2dF result at the 2$\sigma$ level.
While this is an interesting finding it is
clear that the X-ray data is subject to a serious problem with
systematic uncertainties, such as the calibration of the mass-temperature
relation. Therefore the result more likely points to a problem with
the interpretation of the X-ray data than to evidence of a non-zero neutrino
mass.

\begin{figure}[h]
\begin{center}
\vspace*{0.0cm}
\epsfysize=12truecm\epsfbox{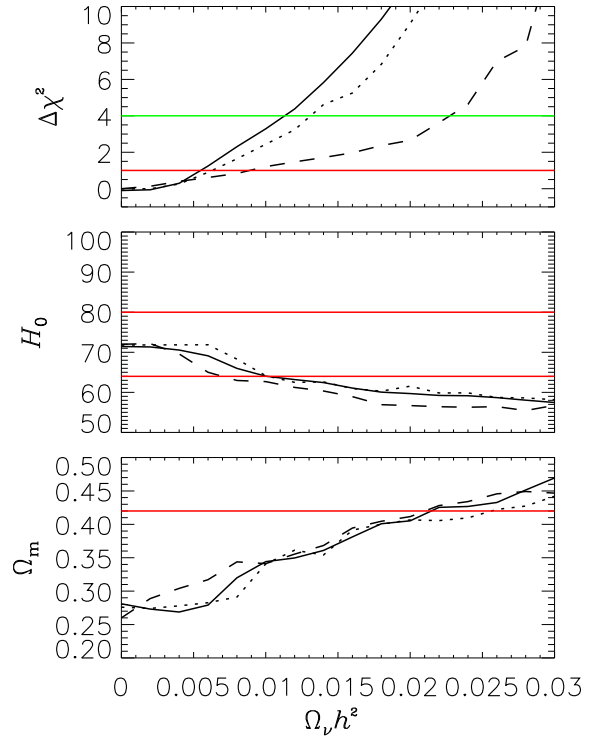}
\end{center}
\vspace*{-1.0cm}
\caption{The top panel shows
$\chi^2$ as a function of $\sum m_\nu$ for different choices
of priors. The dotted line is for WMAP + 2dFGRS data alone,
the dashed line is with the additional Wang {\it et al.} data.
The full line is for additional HST and SNI-a
priors as discussed in the text. The horizontal lines show
$\Delta \chi^2 = 1$ and 4 respectively. 
The middle panel shows the best fit values of $H_0$ for a given
$\sum m_\nu$. The horizontal lines show the HST key project
$1\sigma$ limit of $H_0 = 72\pm 8 \,\, {\rm km}\,{\rm s}\,{\rm Mpc}^{-1}$.
Finally, the lower panel shows best fit values of $\Omega_m$. In this
case the horizontal line corresponds to the SNI-a $1\sigma$ upper limit
of $\Omega_m < 0.42$.}
\label{fig1}
\end{figure}

\section{Sterile neutrinos}
\label{sec:3}

In Ref.~\cite{steen03} it was shown that there is a degeneracy between
the neutrino mass ($\sum m_\nu$) and the relativistic energy density,
parameterized in terms of the effective number of neutrino species, $N_\nu$.

As can be seen from Fig.~2, the best fit actually is actually 
shifted to higher $\sum m_\nu$ when $N_\nu$ increases, and the conclusion
is that a model with high neutrino mass and additional relativistic
energy density can provide acceptable fits to the data. As a function
of $N_\nu$ the upper bound on $\sum m_\nu$ (at 95\% confidence) can
be seen in Table 2.

%
\begin{table}
\caption{95\% C.L. upper limits on $\sum m_\nu$ for different values
of $N_\nu$.}
\label{tab:2}       
\begin{center}
\begin{tabular}{cc}
\hline\noalign{\smallskip}
effective $N_\nu$ &  $\sum m_\nu$ (95\% C.L.) \\
\noalign{\smallskip}\hline\noalign{\smallskip}
3 & 1.01 eV \\
4 & 1.38 eV  \\
5 & 2.12 eV \\
\noalign{\smallskip}\hline
\end{tabular}
\end{center}
\vspace*{1cm}  
\end{table}

This has significant implications for attempts to constrain the
LSND experiment using the present cosmological data.
Pierce and Murayama conclude from the present MAP
limit that the LSND result is excluded \cite{Pierce:2003uh} 
(see also Ref.~\cite{Giunti:2003cf}).

However, for several reasons this conclusion does not follow
trivially from the present data. 
In general the three mass differences implied by Solar, atmospheric
and the LSND neutrino measurements can be arranged into either 
2+2 or 3+1 schemes.
Recent analyses \cite{Maltoni:2002xd}
of experimental data have shown that the 2+2 
models are ruled out. The 3+1 scheme with a single massive
state, $m_4$, which makes up the LSND mass gap, is still
marginally allowed in a few small windows in the 
$(\Delta m^2,\sin^2 2 \theta)$ plane. These gaps are at 
$(\Delta m^2,\sin^2 2 \theta) \simeq  (0.8 \, {\rm eV}^2, 2 \times 10^{-3}),
(1.8 \, {\rm eV}^2, 8 \times 10^{-4}), 
(6 \, {\rm eV}^2, 1.5 \times 10^{-3})$ and
 $(10 \, {\rm eV}^2, 1.5 \times 10^{-3})$.
These four windows corresponds to masses of $0.9, 1.4, 2.5$ and 3.2 eV
respectively.
From the Solar and atmospheric neutrino results the three light
mass eigenstates contribute only about 0.1 eV of mass if they are
hierarchical. This means that the sum of all mass eigenstate is close
to $m_4$.

The limit for $N_\nu = 4$ which corresponds roughly to the LSND
scenario is $\sum m_\nu \leq 1.4$ eV, which still leaves
the lowest of the remaining windows. The second window at $m \sim 1.8$
eV is disfavoured by the data, but not at very high significance.

\begin{figure}[h]
\begin{center}
\vspace*{0.0cm}
\epsfysize=7truecm\epsfbox{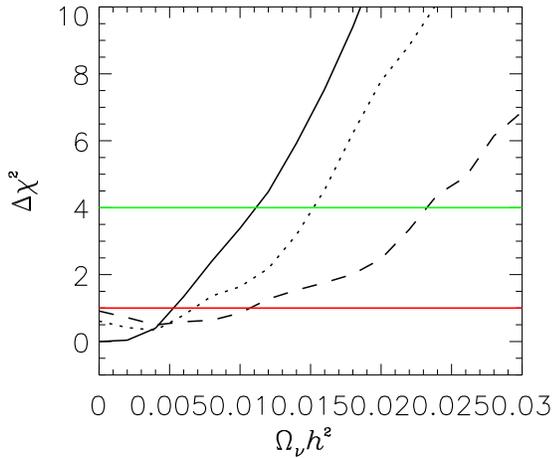}
\end{center}
\vspace*{0.0cm}
\caption{$\Delta \chi^2$ as a function of $\sum m_\nu$ for various
different values of $N_\nu$. The full line is for $N_\nu = 3$, the dotted
for $N_\nu = 4$, and the dashed for $N_\nu = 5$. $\Delta \chi^2$ is calculated
relative to the best fit $N_\nu = 3$ model.}
\label{fig2}
\end{figure}

\section{Discussion}
\label{sec:4}

We have calculated improved constraints on neutrino masses
and the cosmological relativistic
energy density, using the new WMAP data together with data from the
2dFGRS galaxy survey.

Using CMB and LSS data together with a prior from the HST key project
on $H_0$ yielded an upper bound of $\sum m_\nu \leq 1.01$ eV
(95\% conf.). While this excludes most of the parameter range
suggested by the claimed evidence for neutrinoless double
beta decay in the Heidelberg-Moscow experiment, it seems premature
to rule out this claim based on cosmological observations.

Another issue where the cosmological upper bound on neutrino
masses is very important is for the prospects of directly measuring
neutrino masses in tritium endpoint measurements.
The successor to the Mainz experiment, KATRIN, is designed to
measure an electron neutrino mass of roughly 0.2 eV,
or in terms 
of the sum of neutrino mass eigenstates, $\sum m_\nu \leq 0.75$ eV
(see contribution by Guido Drexlin to the present volume).
The WMAP result of $\sum m_\nu \leq 0.7$ eV (95\% conf.) already
seems to exclude a positive measurement of mass in KATRIN.
However, this very tight limit depends on priors, as well
as Ly-$\alpha$ forest data, and the more conservative present
limit of $\sum m_\nu \leq 1.01$ eV (95\% conf.) does not exclude
that KATRIN will detect a neutrino mass.

Finally, we also found that the neutrino mass bound depends on
the total number of light neutrino species. In scenarios with sterile
neutrinos this is an important factor. For instance in 3+1 models
the mass bound increases from 1.0 eV to 1.4 eV, meaning that the
LSND result is not ruled out by cosmological observations yet.

%
%

\end{document}